\documentclass[sigconf, nonacm]{acmart}

\newcommand\vldbdoi{10.14778/3476311.3476346}
\newcommand\vldbpages{2791 - 2794}
\newcommand\vldbvolume{14}
\newcommand\vldbissue{12}
\newcommand\vldbyear{2021}
\newcommand\vldbauthors{\authors}
\newcommand\vldbtitle{\shorttitle} 
\newcommand\vldbavailabilityurl{https://gitlab.com/ViDA-NYU/auctus/auctus}
\newcommand\vldbpagestyle{empty} 

\usepackage{xspace} 
\usepackage{enumitem}
\usepackage{boxedminipage}
\usepackage{subfigure}
\usepackage{hyperref}
\usepackage[a-2b]{pdfx}

\newcommand{\jfcolor}[1]{{\color{blue}#1}} 
\newcommand{\ascolor}[1]{{\color{violet}#1}} 
\newcommand{\abcolor}[1]{{\color{red}#1}} 
\newcommand{\scqcolor}[1]{{\color{olive}#1}} 

\newcommand{\jf}[1]{\jfcolor{\textbf{JF: }\sf #1}}
\newcommand{\as}[1]{\ascolor{\textbf{AS: }\sf #1}}

\newcommand{\ab}[1]{\abcolor{\textbf{AB: }\sf #1}}
\newcommand{\scq}[1]{\scqcolor{\textbf{SC: }\sf #1}}

\newcommand{\prototype}{Auctus}

\newcommand{\systemname}{\texttt{\prototype}\xspace}
\newcommand{\dataset}{dataset\xspace}
\newcommand{\Dataset}{Dataset\xspace}
\newcommand{\datasets}{datasets\xspace}

\newcommand{\inputdata}{$D$\xspace}
\newcommand{\query}{$Q$\xspace}
\newcommand{\outputdata}{$\mathbb{D}$\xspace}

\newcommand{\eg}{e.g.,\xspace}

\newcommand\myparagraph[1]{ \vspace{4pt} \noindent \textbf{#1.}}
\newcommand{\paragraphtt}[1]{\noindent \textit{#1.}}

{\begin{list}{$\bullet$}{%
	\setlength{\labelsep}{2pt}\setlength{\leftmargin}{0pt}%
	\setlength{\labelwidth}{0pt}%
	\setlength{\listparindent}{0pt}}}
{\end{list}}

\usepackage{cuted}
\usepackage{capt-of}

\newcommand{\hide}[1]{}

\begin{document}
\title{\prototype: A Dataset Search Engine for \\ Data Discovery and Augmentation}







\author{Sonia Castelo, R\'{e}mi Rampin, A\'ecio Santos,  Aline Bessa, Fernando Chirigati and Juliana Freire}
\affiliation{%
  \institution{New York University}
}
\email{{s.castelo, remi.rampin,  aecio.santos, aline.bessa, fchirigati, juliana.freire}@nyu.edu}

\begin{abstract}
The large volumes of structured data currently available, from Web
tables to open-data portals and enterprise data, 
open up new opportunities for progress in answering many important scientific, societal, and business questions. However,  finding 
relevant data is difficult. While search engines have addressed this problem for Web documents, there are many new challenges involved in supporting the discovery of structured data.
We demonstrate how the \systemname dataset search engine addresses some of these challenges.
We describe the system architecture and  how users can explore datasets through a rich set of queries. We also present case studies which show how \systemname supports data augmentation to improve machine learning models as well as to enrich analytics.
%

\end{abstract}

\maketitle

\pagestyle{\vldbpagestyle}
\begingroup\small\noindent\raggedright\textbf{PVLDB Reference Format:}\\
\vldbauthors. \vldbtitle. PVLDB, \vldbvolume(\vldbissue): \vldbpages, \vldbyear.\\
\href{https://doi.org/\vldbdoi}{doi:\vldbdoi}
\endgroup
\begingroup
\renewcommand\thefootnote{}\footnote{\noindent
This work is licensed under the Creative Commons BY-NC-ND 4.0 International License. Visit \url{https://creativecommons.org/licenses/by-nc-nd/4.0/} to view a copy of this license. For any use beyond those covered by this license, obtain permission by emailing \href{mailto:info@vldb.org}{info@vldb.org}. Copyright is held by the owner/author(s). Publication rights licensed to the VLDB Endowment. \\
\raggedright Proceedings of the VLDB Endowment, Vol. \vldbvolume, No. \vldbissue\ %
ISSN 2150-8097. \\
\href{https://doi.org/\vldbdoi}{doi:\vldbdoi} \\
}\addtocounter{footnote}{-1}\endgroup


\section{Introduction}
\label{sec:intro}

With the push towards transparency and openness, scientists, governments, and companies have been increasingly publishing structured
datasets on the Web. Google Dataset Search alone indexes over 30 million datasets~\cite{google-dataset-search@ISWC2020}.
The availability of these data 
%
creates opportunities to answer many important scientific, societal, and business questions.

\myparagraph{The Need for Data Discovery}
%
While data are abundant, finding \textit{relevant data} is difficult. 
Data are spread over a large number
of sites and repositories. Recognizing this challenge, a number of approaches have been proposed to \emph{organize and index data collections}~\cite{chapman@2019}, from domain-specific repositories such as NYCOpenData, which collects datasets from the various NYC agencies~\cite{nycopendata}, general data portal infrastructure and data lakes~\cite{socrata,dataverse}, to Google Dataset Search, which indexes a wide range of datasets published on the Web~\cite{noy@www2019}. 
%
%
While these present a significant step towards simplifying data discovery, they have an important limitation: they only support simple, keyword-based  search queries over published dataset metadata. This greatly limits a user's ability to express information needs. In addition, published metadata is often incomplete, and in some cases it can be inconsistent with the actual data. Thus, relying solely on the metadata also limits the discoverability of datasets.

\begin{figure}
\centering
\includegraphics[width=0.39\textwidth]{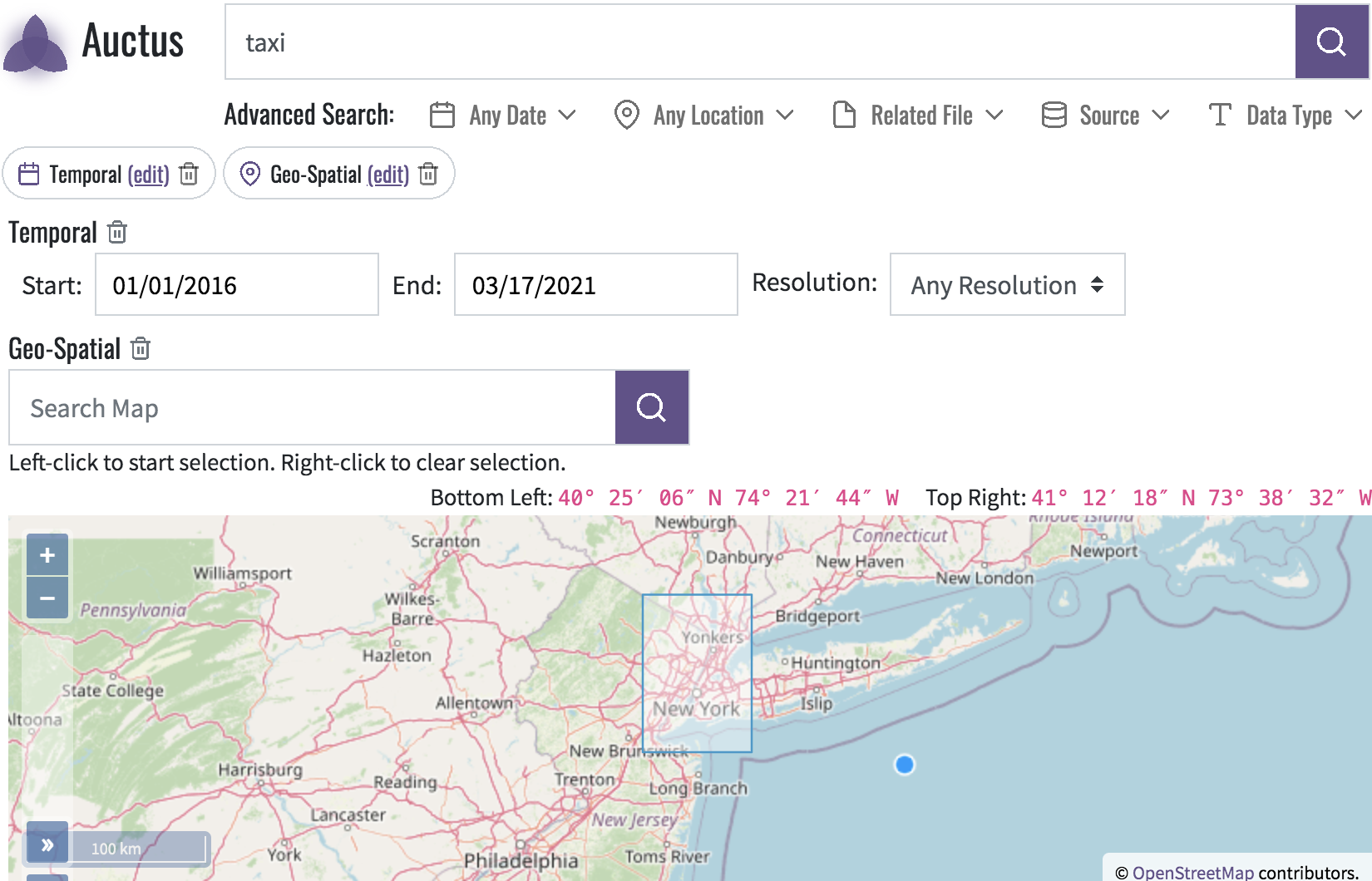}
\vspace{-.3cm}
\captionof{figure}{Searching for datasets that mention  \emph{``taxi"} and contain records within the NYC area for the 2016-2021 period.
\label{fig:actus-query}}
\vspace{-1.1cm}
\end{figure}

\myparagraph{Our Approach}
%
\systemname is an  \textit{open-source \dataset search engine that was designed to support data discovery and augmentation}. 
It  supports a rich set of discovery queries: in addition to keyword-based search, users can specify spatial and temporal queries, data integration queries (i.e., searching for datasets that can be concatenated to or joined with a query dataset), and they can also pose complex queries that combine multiple constraints, as shown in Figure~\ref{fig:actus-query}. 
%
These queries are enabled in part by 
a data profiler~\cite{datamart-profiler}
that we developed to extract useful information from the actual datasets. This includes not only summaries (or sketches) of column contents, but also their data types. In particular, it detects columns that contain spatial and temporal information. This information is then used to construct indices that support efficient query evaluation. 
Users can explore large dataset collections  through an easy-to-use interface that guides them in the process of specifying complex queries. To help users identify relevant datasets, \systemname displays snippets that summarize the contents of datasets.


%
\systemname has been developed in the context of the DARPA D3M program~\cite{d3m}, and it is currently being used in production and to support research of different groups in the project~\cite{santos2019visus, hong2020towards, chepurko2020arda}. 
In this demo paper, we give an overview of the architecture and features of \systemname (Section~\ref{sec:auctus_system}), and 
 discuss a few use cases that we will present during our demonstration  (Section~\ref{sec:use_cases}).
Demo visitors will  be able to interact directly with \systemname to query over 19,000 \datasets.
%



\hide{\jf{when including references, if a paper has been published, use the published version. Here, the ref to ARDA is the arxiv version, not the published version. We also should make all references consistent, e.g., use the same title for conferences.}
\scq{will work on references}}

\myparagraph{Related Work} Our work\hide{builds on a large body of research that includes} is related to  methods for data profiling~\cite{abedjan2015profiling}, and for the  discovery of joinable~\cite{fernandez@icde2019} and unionable~\cite{nargesian2018table} tables --- which we use in our system. 
Moreover, \systemname is related to multiple data discovery systems~\cite{zhu2017interactive, fernandez@icde2018}. \hide{~\cite{deng2017data, zhu2017interactive, fernandez@icde2018} have been proposed.}
In particular, it is related to \hide{For instance, Data Civilizer~\cite{deng2017data} and Aurum~\cite{fernandez@icde2018} use a linkage graphs model to store relationships to support the identification of relevant data to a user task.} 
JUNEAU~\cite{zhang2019juneau}\hide{\scq{lack of space},zhang2020finding}, a system that supports search capabilities in data science environments such as Jupyter notebooks. \hide{ Toronto Open Data Search~\cite{zhu2017interactive} allows navigation of joinable tables from open data portals.} 
While our work shares some features and goals with these systems, 
\systemname provides a unique search interface that combines the familiar user interface (UI) of Web search engines with visual metaphors of modern data analytics tools such as Tableau~\cite{tableau}. 
\systemname' UI allows users not only to perform expressive queries that include multiple constraints,  but also allows users to make sense of the tables returned as search results. Further, it supports the  materialization of joins and unions into CSV and D3M file formats,\hide{dataset format (\texttt{application/vnd.d3m-dataset}),} and  communication through a REST API that allows for integration  \hide{, which allowed it to be integrated} with external interactive data analytics systems~\cite{santos2019visus, bessa@sigmod2021, shang2019democratizing}.

\section{The Auctus System}
\label{sec:auctus_system}

\subsection{Auctus Architecture}
\label{subsec: auctus_arquitecture}

\begin{figure}
\centering
\includegraphics[width=0.34\textwidth]{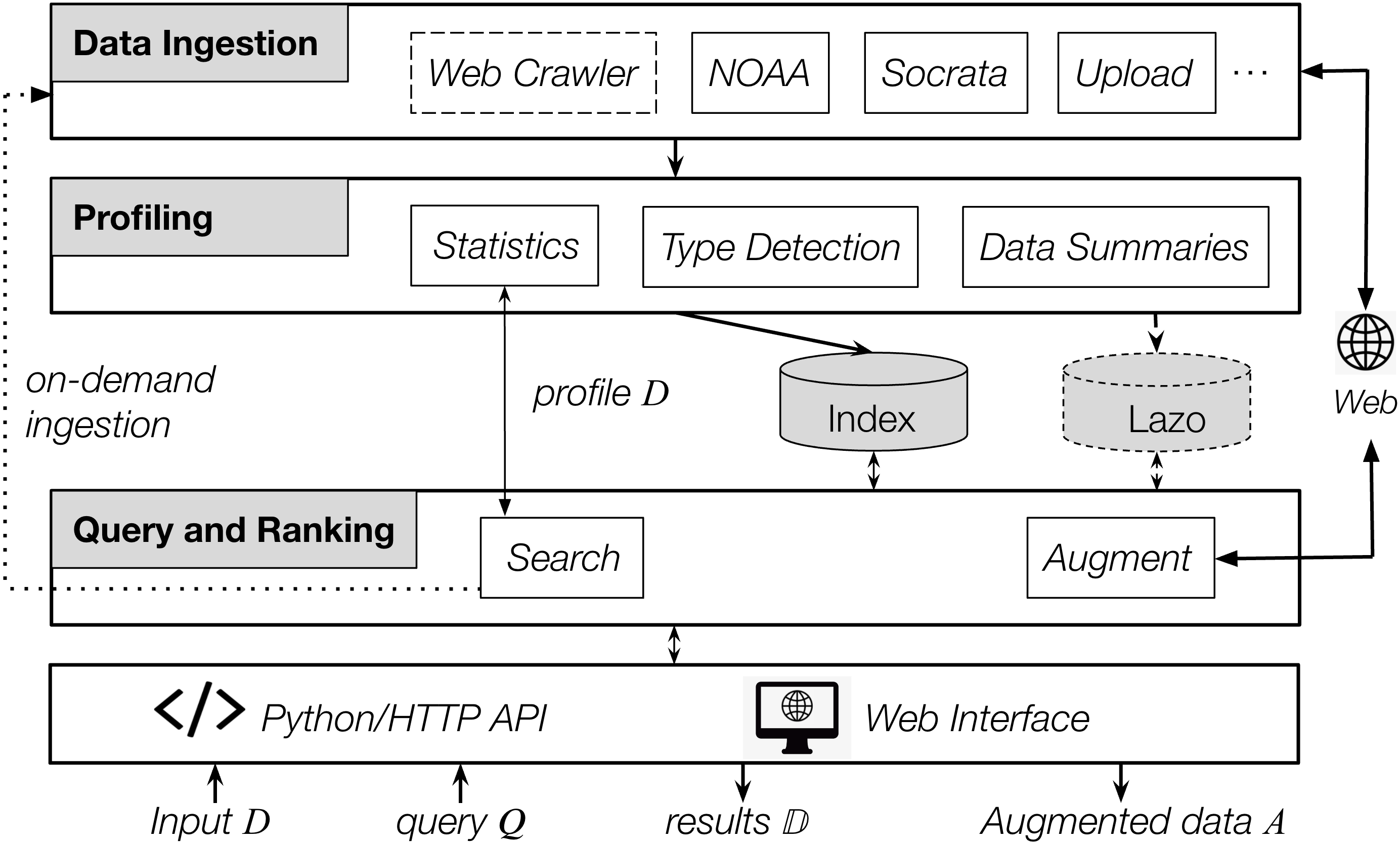}
\vspace{-.2cm}
\captionof{figure}{Overview of \systemname' architecture.
\label{fig:architecture}}
\vspace{-.5cm}
\end{figure}

The high-level architecture of \systemname  is depicted in Figure~\ref{fig:architecture}. In what follows, we describe its key components.

\myparagraph{Data Ingestion}
%
\systemname makes use of plugins to retrieve \datasets
from \hide{specific} repositories using their \hide{respective} APIs.
This makes the system extensible and able to ingest data from many different sources.
Currently, it supports 
Socrata~\cite{socrata} (a platform for open government data),
Zenodo~\cite{zenodo} (open-access data repository),
the World Bank Open Data~\cite{worldbank} (datasets on global development),
among others.
%
%
%
\hide{While our approach is different from Google \Dataset Search~\cite{noy@www2019}, we are currently working on leveraging multiple search engines (including Google \Dataset Search) to discover new \datasets that are not present in \systemname index. These search engines will \textit{complement} our current approach. \ab{commenting for lack of space, and because it's not implemented yet}
}
\systemname also allows users to upload their own
\datasets.

\myparagraph{Profiling}
Once \datasets are ingested,
\systemname profiles them  to infer relevant metadata necessary to support discovery queries as well as to construct dataset summaries
for result presentation.
%
%
The profiler performs different tasks, including:
\textit{type detection}, i.e., it detects whether columns correspond to   categorical, numerical, spatial, or temporal attributes; 
type-dependent \textit{statistics computation}  (\eg frequency of  values, mean, and variance for numerical values),  
and \textit{data summarization} (see below). 

\myparagraph{Storing Data and Data Summaries}
Since \systemname was designed to serve as a dataset discovery system, 
it stores dataset summaries instead of the full datasets\hide{instead of storing the full datasets, it stores data summaries}. These summaries are concise and  sufficient to create the indices required to answer all queries supported by the system.
%
Currently, \systemname creates summaries
for categorical, numerical, spatial, and temporal
attributes. 
The data summaries generated by \systemname are represented
by the ranges of their corresponding attributes.
During the search phase, \systemname uses these summaries to estimate
the size of the intersection between two attributes, concluding 
whether a join is feasible.
To estimate join intersections well, \systemname captures  fine-grained ranges for data attributes by using the clustering algorithm $k$-means. 
This strategy produces the desired results while also being efficient.
\systemname also stores provenance information to enable the retrieval of the datasets. This allows the system to perform augmentations, and users to download the datasets. \systemname can also cache datasets for efficiency purposes. 
%

%
%
%

\myparagraph{Indices}
After the metadata are generated, including data summaries,
they are indexed in an Elasticsearch~\cite{elasticsearch} server.
Numerical and temporal summaries are indexed using
range data types, and spatial summaries are indexed
using geo-shape data types.
To support joinable dataset search, we use Lazo~\cite{fernandez@icde2019}, which is a method for set-overlap search based on MinHash sketches and locality-sensitive hashing (LSH), to build an index for categorical attributes. 
%

\hide{\jf{the text in this paragraph overlaps with the description we have in 2.2; maybe we should have a high-level overview here -- at the level of the architecture and how this connects to the other components, and leave the details of specific queries to 2.2. What do people think?}
\scq{I don't think we should merge them. In Section 2.2, we just describe the user interface, here (section 2.1) we are explaining how these processes happen internally.} \ab{I agree, Sonia} \as{I think this should actually get more technical about query processing. I tried to add more details.}}

\myparagraph{Querying and Ranking}
%
\systemname supports queries that combine multiple constraints including keywords, temporal, spatial, data type, and data source.  The system also supports \emph{data integration queries}: Given an input dataset $D_Q$, \systemname allows the user to search for datasets that can be concatenated to or joined with $D_Q$.
%
%
%
The \dataset $D_Q$ must first be uploaded using a provided API or selected from the set of indexed datasets. The system will then generate a dataset profile which is used to probe those indices that support join and union. 
Finally, the lists of matching datasets from different indices are merged, ranked, and returned as  search results. 


\hide{
\jf{Merge the join and union search with Secion 2.2} \scq{I dont think we should merge because of the same reasons that I mentioned above. Maybe reduce it further.}
}
\paragraphtt{Join Search}
\hide{\jf{We should say that we use Lazo} \scq{we already mention that when we talked about 'Indices' .} \as{We don't use only Lazo. I tried to add more details, but the text is a bit involved.}} 
To find other \datasets that can be joined with \inputdata,
\systemname first searches, for each attribute $a$ of \inputdata,
which other attributes, in the index that corresponds to $a$'s data type (e.g., temporal attributes searched in the Elasticsearch range index, or categorical attributes probed against the Lazo index), have summaries 
intersecting the summary of $a$.  
Every \dataset that has at least one intersecting attribute is 
a potential join result. 

\paragraphtt{Union Search}
To find  \datasets that can be concatenated with \inputdata,
the indices are searched for any \dataset that has attributes with the same data types present in  
\inputdata, as well as similar names.
Name similarity is computed with \hide{accomplished by using}
the \textit{fuzzy query}
in Elasticsearch. \hide{, which applies Levenshtein distance for the similarity search.} 
\hide{The union search does not require all of the attributes from \inputdata
to be matched: a result \dataset may only match a subset of \inputdata.}
Results from union searches may match only a subset of \inputdata's attributes. 

\paragraphtt{Ranking}
The results of join and union searches, after being filtered  based on query \query, are ranked and returned as \outputdata. 
Join results are ranked based on the intersection of the summaries; \hide{:
\datasets with higher intersection are ranked higher.}
union results are ranked based on the Levenshtein similarity between the names of the matching attributes. 

\hide{
\jf{we should merge this with section 2.2} \scq{I dont think we should merge because of the same reasons that I mentioned above.}
}
\myparagraph{Augmentation}
Besides providing search capabilities for data augmentation, \systemname  also performs the actual augmentation. 
Users can choose a \dataset $R\in$ \outputdata, and \systemname will materialize it (using the provenance annotated
in the metadata) and perform the join or union operation with \inputdata, returning the new, augmented \dataset $A$.
If multiple attribute pairs match between $R$ and \inputdata for a join operation, users can choose which pair(s) they want for the join. For temporal and spatial joins, the attributes are translated into the same resolution before the augmentation.

\vspace{-.3cm}
\subsection{Auctus User Interface}
\label{subsec:auctus_interface}

\systemname provides an easy-to-use interface where users can query for datasets, explore search results including data exploration and augmentation options, explore ingested datasets, and  upload new datasets.

\begin{figure}
\centering
\includegraphics[width=0.44\textwidth]{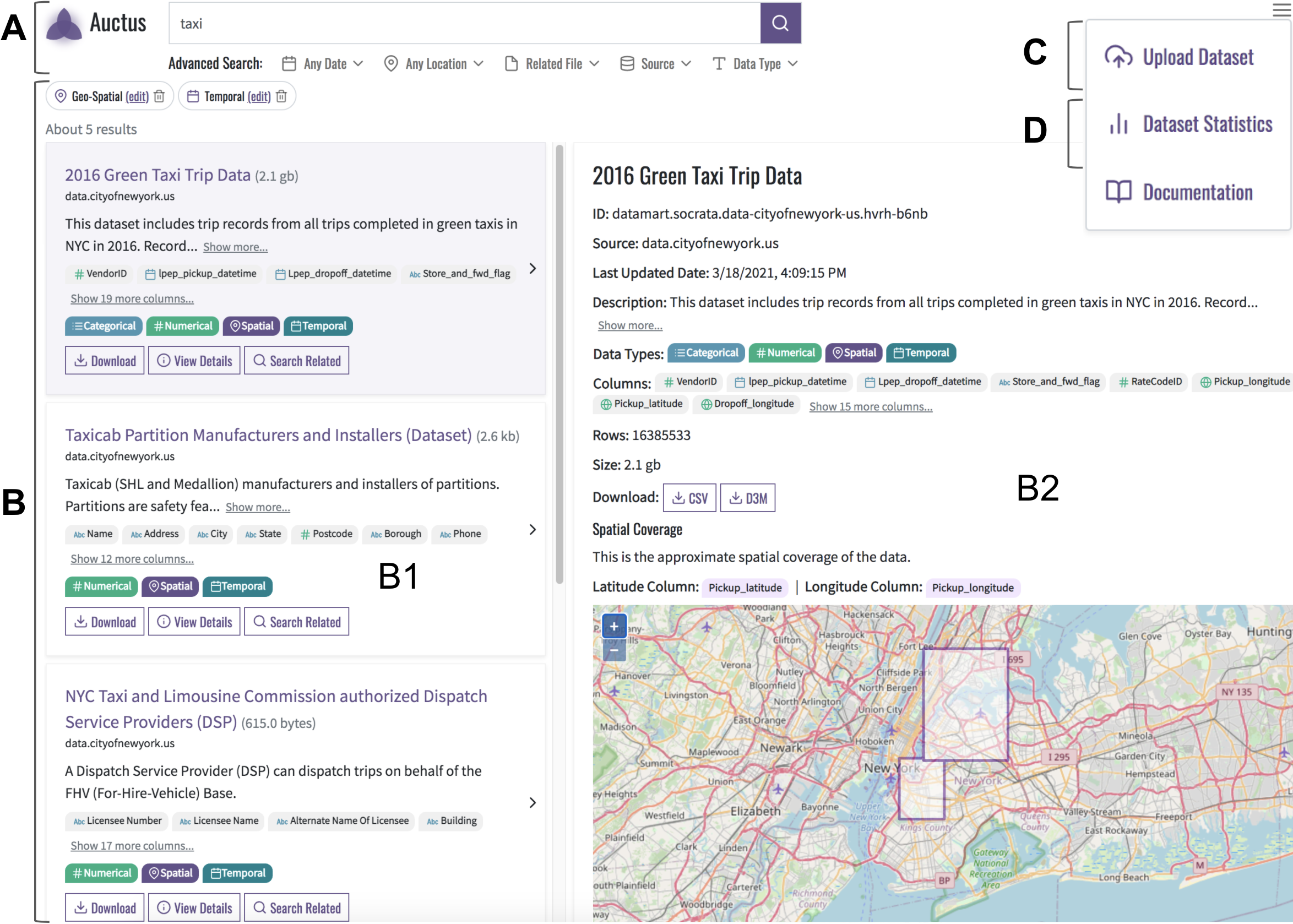}
\vspace{-.1cm}
\captionof{figure}{Components of \systemname' user interface:    keyword and filter-based search box \textbf{(A)}; search results  \textbf{(B)}; dataset-snippets \textbf{(B1)}; dataset summary \textbf{(B2)};  dataset upload \textbf{(C)};  dataset collection statistics  \textbf{(D)}.
\label{fig:teaser}}
\vspace{-.7cm}
\end{figure}

\begin{figure*}
    \vspace{-.3cm}
    \centering
    \subfigure[]{\includegraphics[width=0.28\textwidth]{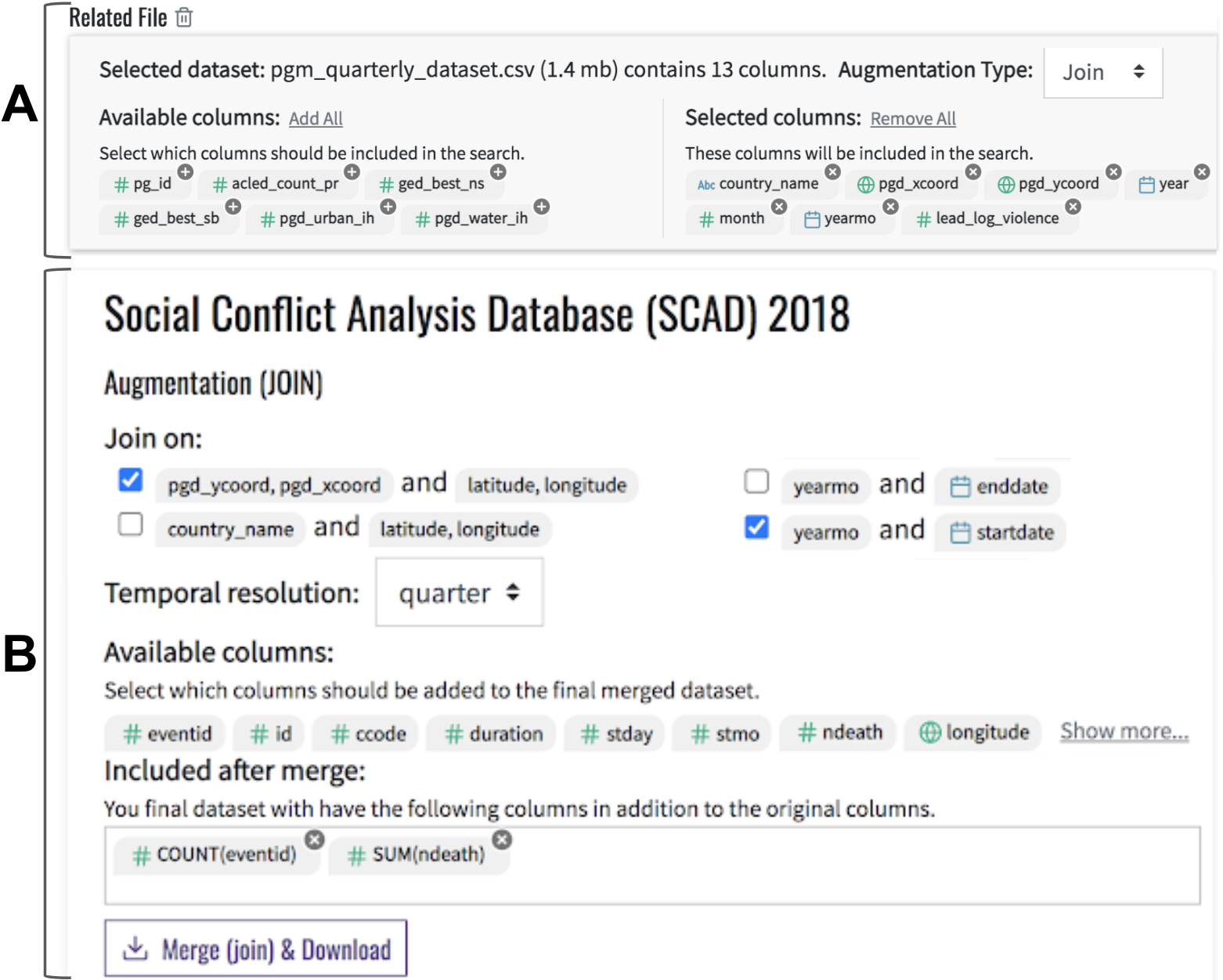}\label{fig:augmentation_interface}} 
    \subfigure[]{\includegraphics[width=0.3\textwidth]{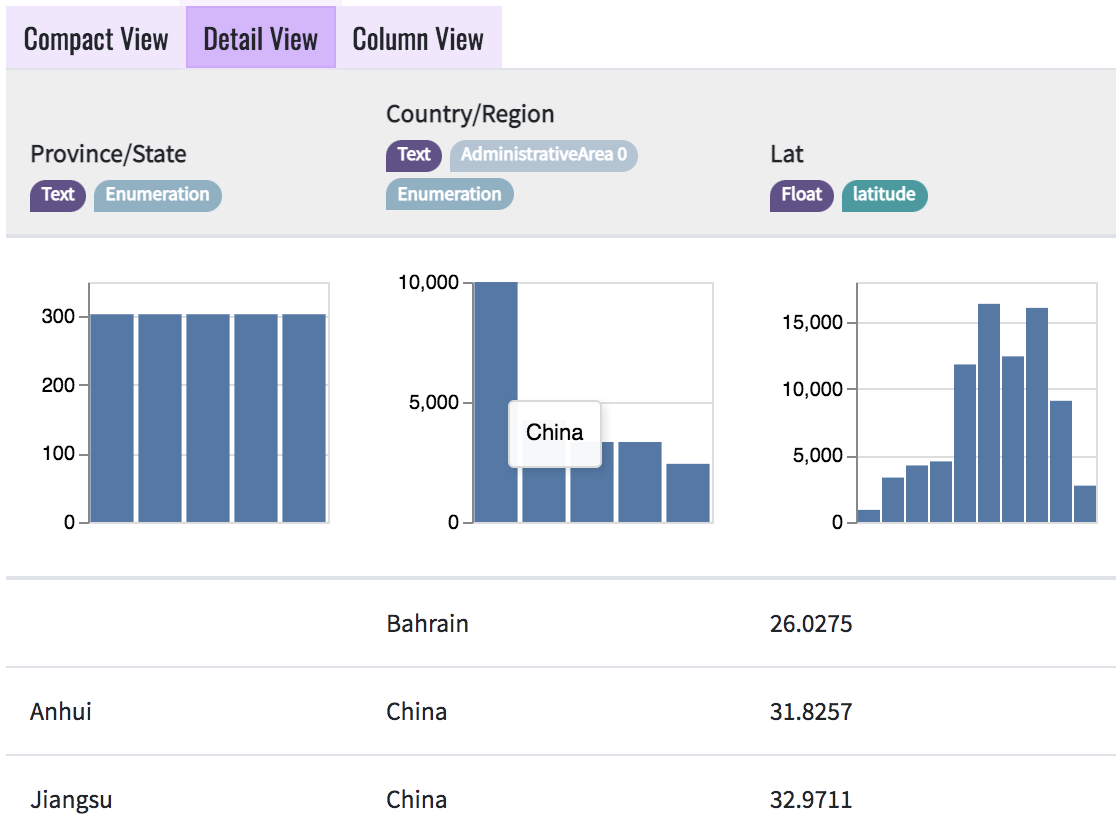}\label{fig:data_summary_views}} 
    \subfigure[]{\includegraphics[width=0.36\textwidth]{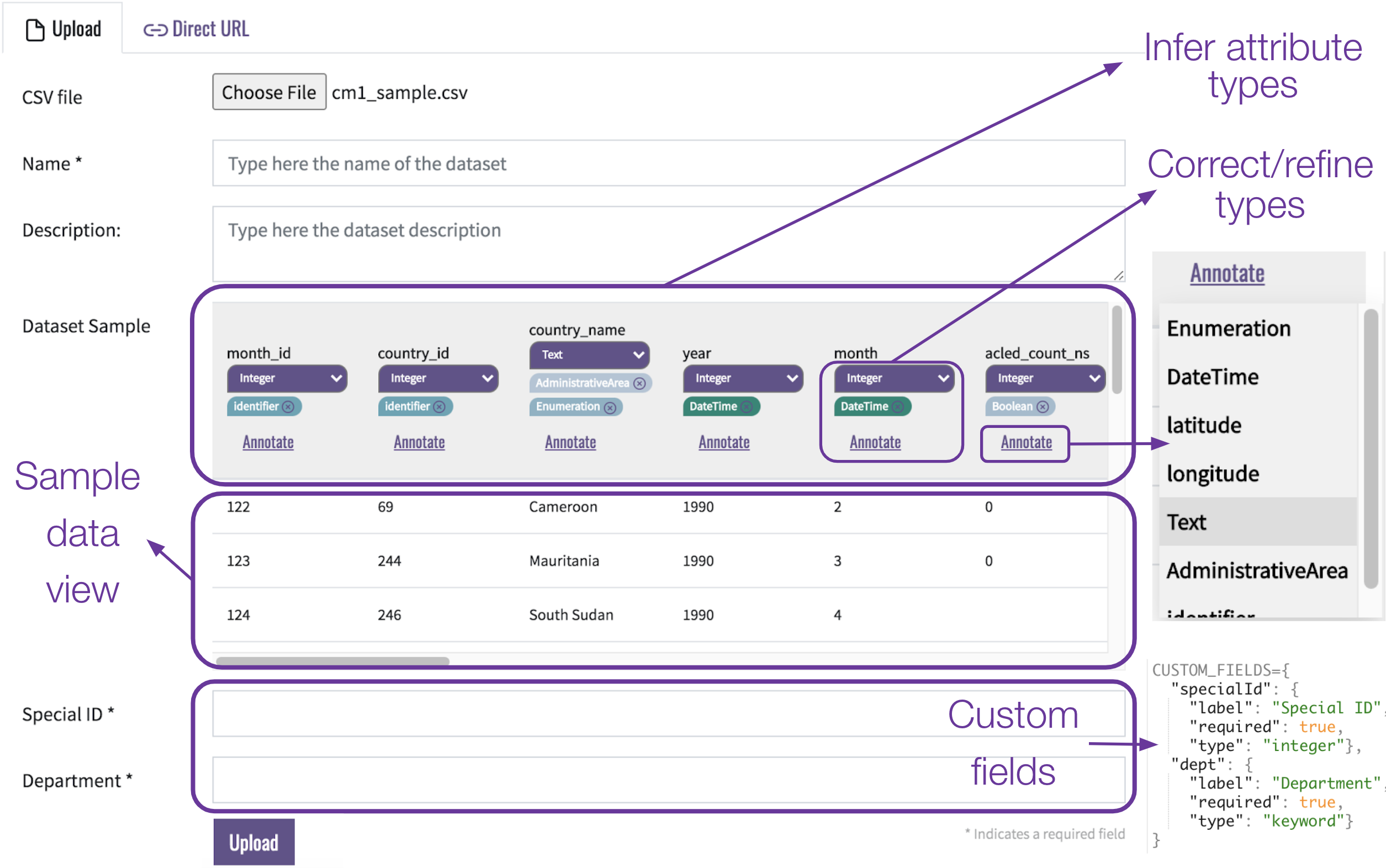}\label{fig:upload_page}} 
    \vspace{-.34cm}
    \caption{(a) A data integration query in [A], and the augmentation interface in [B]. (b) Data Summary views. (c) Upload page.}
    \label{fig:filters_datasummary_upload}
\end{figure*}

\myparagraph{Data Discovery Queries}
Users can query indices \hide{for datasets} by specifying keywords and constraints using various filters (see Figure~\ref{fig:teaser}(A)).
\hide{
This page allows the user to query the index for datasets (See figure \ref{fig:teaser}(A)). They can enter keywords in the search box and also add  constraints based on filters, explained as follows. \hide{on the dates and times present in resulting datasets. Users can also select specific sources to only retrieve datasets obtained from them. 
If the user provides data of their own, the system will try to find datasets that can be joined or unioned with their data. }
}

\paragraphtt{Temporal and Spatial Search}
Users search for datasets by specifying a date range (Figure~\ref{fig:actus-query}) -- datasets containing a temporal column that overlaps with that range will be retrieved. They can \hide{further} refine the search by specifying a \hide{desired} temporal resolution (e.g., year). 
To perform a spatial search, users can either draw a bounding box around a geographical area on the map (Figure~\ref{fig:actus-query}), or specify an administrative area, which \systemname translates into a polygon. Datasets containing spatial attributes that overlap with the search polygon are returned.
%

\paragraphtt{Source filter}
The source filter allows users to restrict the data sources of interest, and only results from the selected sources are retrieved. 

\hide{
\ab{where is the data type filter?}
\scq{done}
}
\paragraphtt{Data Type filter}
The data type filter allows users to search for datasets based on the types of their attributes, e.g., categorical, numerical, spatial and temporal, which were inferred by the profiler. 

\paragraphtt{Data Integration Queries}
The related file filter allows users to find datasets that can augment a given input dataset (see Figure~\ref{fig:augmentation_interface}(A)). The user can upload the input dataset or select a dataset from a set of search results.
Result snippets for data integration queries include an “augment options” button\hide{(Figure~\ref{fig:filters_datasummary_upload}(a))}, that allows users to request and customize the augmentation to be performed by \systemname.

\myparagraph{Result Presentation and Exploration}
Unlike Web documents which can be summarized using short text snippets, datasets have many different facets that the user must consider to determine their relevance. Thus, an important challenge for Auctus is how to present search results.
Figure~\ref{fig:teaser} shows the results for the query specified in Figure~\ref{fig:actus-query}.
%
%
On the left, there is a list of search results displayed as snippets (Figure \ref{fig:teaser}(B1)). Users can select a dataset to inspect its details (Figure \ref{fig:teaser}(B2)), which include  description, source, attribute names and types, as well as a summary of the dataset's contents. \hide{As illustrated in this figure, for} For  spatial datasets, a visualization showing the geographical extent  covered by the dataset \hide{records} is displayed. 
%
The summary also includes a sample of the dataset records and statistics about its  columns (Figure~\ref{fig:filters_datasummary_upload}(b)). 
The tabs above the dataset sample allow for the \hide{selection of different summary views which show}visualization of these statistics in different levels of detail. For example, \emph{Detail View} shows \hide{plots of} the columns' value distributions. 

\paragraphtt{Augmentation}
If the user provides an input file through the Related File filter, the snippets  \hide{also} display a button “Augment Options” under each \hide{search} result on the left. If the user clicks on this button, the augmentation interface will be shown (see Figure~\ref{fig:augmentation_interface}(B)). 
If the \hide{search} result is a union, users can select matched pairs of columns, and  the columns from the result (right) will be appended to the matching columns of the input data (left).
If the result is a join (an example is shown in Figure~\ref{fig:augmentation_interface}(B)), users can select the columns to be matched at the top of the augmentation screen. Those are the “join keys” that determine which records from the input data should be matched with records from the selected  result. Under this area, there is an interface  to include columns from the selected  result into the augmented table. Users can   grab the desired columns from the “available columns” area, drag them to the “included after merge” box, and drop them over the aggregation function they wish to use for that column.

\hide{
\ab{I think we should use uploaded instead of ingested as often as possible. the interface talks about upload a lot.} 
\scq{"ingested" includes both the ones that have been uploaded by the users and the ones discovered by Auctus. At the end both are indexed/ingested by Auctus. I clarified in the intro of Section \ref{sec:auctus_system} what means "ingested datasets" }
}

\myparagraph{Uploading Data and Curating Metadata}
Users can add new datasets to \systemname. 
After loading a data file, \systemname automatically infers its  data types with our datamart-profiler library~\cite{datamart-profiler}. As any \hide{automated} method for type inference, our profiler is not fool-proof and can derive incorrect results. To address that, \systemname enables users to correct data types manually, and to provide additional annotations for the columns. \systemname also displays a dataset sample so that the user can verify if the detected data types  are correct, and check the uploaded data. 
Additionally, \systemname provides support for custom metadata fields (e.g.,  data source or grid size). Since these fields can vary widely for different applications, we defined a flexible configuration schema that allows users to customize them for different deployments. 
The upload page is shown in Figure~\ref{fig:upload_page}. 

\subsection{Scalability and APIs}

\systemname has been implemented with scalability in mind:
the search engine is entirely containerized using Docker~\cite{docker}. 
Each data discovery plugin\hide{, for instance,} is an independent container,
allowing multiple plugins to be executed in parallel.
\systemname can also spin up as many profiling and query
containers as required in response to load.
Our system can be accessed via a Web UI
\cite{auctus-interface}
or programmatically via Python and REST APIs. 
So far, we have indexed over 19,000 \datasets.
\section{Use Cases}
\label{sec:use_cases}
Visitors will be able to see how \systemname is used for \hide{three} different use cases  (presented below), and interact with it through its interface. 

\myparagraph{Bicycle Usage Prediction}
Predicting the number of daily bicycle trips is an important step 
towards implementing better policies for this means of transportation in NYC.  
%
Consider that an expert from the NYC department of transportation  decides
to build a model using the East River Bicycle Trips data.
This department installed automated counters in all East River bridges, which 
provide the number of bikes crossing them on a daily basis. This number can 
be used as a proxy for the overall bicycle usage in NYC. 
Besides daily bicycle counts, the data also contains maximum daily temperatures for NYC. 
Unfortunately, her initial dataset only covers bicycle usage for April 2020, which is not very informative. 
To take into account a larger period of time, she uses \systemname 
to find compatible data for more months of 2020, which can then be 
 concatenated to her original input data within our system. 
When using a random forest regressor to predict the number of bicycle trips using daily temperatures as a model feature, she obtains an 
$R^2$ score close to 0.25. She is relying on a single feature to make her predictions, so her model is probably underfitting. 
To improve it, she uses \systemname again to find and augment her current input data with weather information, including daily rainfall levels. The $R^2$ score then increases to approximately 0.56, 
which represents a substantial improvement. A video demonstrating this use case is available at \url{http://bit.ly/auctus-demo}.

\hide{\myparagraph{Vision Zero}
As part of the NYC Vision Zero
Initiative~\cite{visionzero},
an expert from the Department of Transportation is trying to devise policies
to reduce the number of traffic fatalities and improve safety in NYC streets.
Initially, she uses data about traffic collisions and taxi trips to create a 
model that predicts the number of collisions, allowing her  
 to explore \textit{what-if} scenarios.
To improve the model, she looks for weather information,
as weather conditions can potentially increase the number of collisions.
She uses \systemname to find \datasets that
have the keyword ``weather'', and that are related to her input data.
Because many collisions also involve cyclists,
particularly after the arrival of the Citi Bike in NYC,
she also searches 
for \datasets 
having the keywords ``citi bike''.
After doing the augmentations with weather and Citi Bike data,
the model achieves a good performance for her analysis.}
\hide{
\ab{articulate the challenges Vito faced;  explain
how auctus greatly simplifies both the discovery and integration tasks he must perform.}
\scq{Done}
}
\myparagraph{Conflict Forecasting}
While analyzing conflict forecasting problems, researchers often  use predictive modeling to guide policy-making decisions, and to assess and compare theories of conflict~\cite{d2020conflict}. In this context, it is crucial to identify new data sources, merge those data, and evaluate the contribution of different features.
Furthermore, most conflicts materialize as events,  and if they occur in heterogenous spatio-temporal levels and need to be analyzed in tandem, data integration can be challenging. Consider that a researcher is studying conflict events in Africa. She uses the 
grid ~\cite{prio-grid}
dataset for Africa with conflict events aggregated into quarterly counts\hide{ from 2010 to 2020} to predict state-based conflicts.
%
%
Since protest events data can be very useful, and since her grid dataset does not have any measure of protests, she decides to use \systemname to discover and join this type of dataset to her input data. She uses "protest"  as a keyword, and also uploads the grid dataset as an input query in \systemname. 
After running the query, the highest ranked result is the Social Conflict Analysis Database (SCAD). 
To verify its suitability, she explores the dataset through the "View Detail" tab.
After inspecting the Summary Data view, she realizes that SCAD  contains information not only on protests, but also on other destabilization events. 
She can also quickly see that it captures events in Africa because of the map visualizations, and also finds spatio-temporal matches. 
She then checks the possible augmentation options moving in to the augmentation interface.  As shown in in Figure~\ref{fig:augmentation_interface}(B), \systemname automatically detects joinable columns. She then selects the temporal and spatial levels of her interest, and uses aggregation functions to count the number of destabilizing events and to sum the number of  fatalities. Next, she presses the augmentation button and all the events in SCAD become aggregated into the grid dataset.  
The augmented dataset can be used to explore new research questions, such as whether the number of destabilizing events at time \textit{t} is a predictor for state-based violence at time \textit{t+1}. 


\begin{acks}
This work was partially supported by the
DARPA D3M program and NSF award OAC-1640864. Any opinions,
findings, and conclusions or recommendations expressed in this
material are those of the authors and do not necessarily reflect the
views of NSF and DARPA.
\end{acks}

\bibliographystyle{ACM-Reference-Format}
\bibliography{paper}

\end{document}